\documentclass[aps,prc,twocolumn,superscriptaddress,showpacs,floatfix,nobibnotes]{revtex4}
\usepackage{epsfig}
\usepackage{epstopdf}
\usepackage{color}

\usepackage{graphicx}
\usepackage{longtable}
\usepackage{CJK}

\usepackage{graphicx}
\usepackage{longtable}
\usepackage{CJK}
\usepackage{color}

\usepackage{mathptmx, courier, pifont}
\usepackage[scaled=0.92]{helvet}
\usepackage[T1]{fontenc}
\usepackage{textcomp}
\usepackage{fancyhdr}

\begin{document}


\begin{CJK*}{GB}{}

\title{Triaxial dynamics in the quadrupole-deformed rotor}

\author{Qiu-Yue Li}
\affiliation{Department of Physics, Liaoning Normal University,
Dalian 116029, China}

\author{Xiao-Xia Wang}
\affiliation{Department of Physics, Liaoning Normal University,
Dalian 116029, China}

\author{Yan Zuo}
\affiliation{Department of Physics, Liaoning Normal University,
Dalian 116029, China}

\author{Yu Zhang*}
\affiliation{Department of Physics, Liaoning Normal University,
Dalian 116029, China}

\author{Feng Pan}
\affiliation{Department of Physics, Liaoning Normal University,
Dalian 116029, China}

\date{\today}

\begin{abstract}
The triaxial dynamics of the quadrupole-deformed rotor model of both
the rigid and the irrotational type have been investigated in
detail. The results indicate that level patterns and E2 transitional
characters of the two types of the model can be matched with each
other to the leading order of the deformation parameter $\beta$.
Especially, it is found that the dynamical structure of the
irrotational type with most triaxial deformation ($\gamma=30^\circ$)
is equivalent to that of the rigid type with oblate deformation
($\gamma=60^\circ$), and the associated spectrum can be classified
into the standard rotational bands obeying the rotational
$L(L+1)$-law or regrouped into a new ground- and $\gamma$-band with
odd-even staggering in the new $\gamma$-band commonly recognized as
a signature of the triaxiality. The differences between the two
types of the model in this case are emphasized especially on the E2
transitional characters.

\end{abstract}


\maketitle

\thispagestyle{fancy} \fancyhead{} \lhead{Submitted to Chinese
Physics C} \chead{} \rhead{} \lfoot{*dlzhangyu$_-$physics@163.com}
\cfoot{\thepage} \rfoot{}
\renewcommand{\headrulewidth}{0pt}
\renewcommand{\footrulewidth}{0.7pt}

\end{CJK*}

\begin{center}
\textbf{I Introduction}
\end{center}

The quantum rotor has been widely applied to explain rotational
excitations in molecules and
nuclei~\cite{Kronig1937,Casimir1931,Bohr1952,Davydov1958}. For heavy
or medium mass nuclei, it is often assumed that there is a
quadrupole-deformed surface of these ellipsoidal
nuclei~\cite{Bohrbook,Greinerbook,Ui1970}. Thus, rotational
excitations in these nuclei may be schematically described by
rotational dynamics of an ellipsoid with quadrupole deformation.

Although the quantum rotor is illustrated in many
textbooks~\cite{Bohrbook,Greinerbook}, a detailed comparison of
rotational dynamics of different types of quadrupole-deformed
rotor~\cite{Ui1970} is still absent, which, however, should be made,
especially when the rotor is applied to describe rotational motions
in deformed nuclei. Specifically, the triaxial
rotor~\cite{Davydov1958}, which has been widely using as a basic and
simple description of nuclear collectivity, has been realized
microscopically within the SU(3) shell
model~\cite{Leschber1987,CDL1988,Naqvi1995} and algebraically in the
interacting boson model~\cite{Smirnov2000,Thiamova2010,Zhang2014}.
Particularly, our recent analysis~\cite{Zhang2014} shows that the
$E2$ properties in the SU(3) image of the quadrupole-deformed rotor
is closer to those obtained from the rgid type rotor. On the other
hand, the values of the moment of inertias extracted from
experiments may approach to those obtained from the irrotational
type rotor. The dynamical differences between the two types of rotor
in an axial-deformed case is well
known~\cite{Bohrbook,Greinerbook,Ui1970}. However, the situations in
the triaxial-deformed case remain to be investigated. More recently,
a triaxial rotor model with independent inertia and $E2$ tensors was
suggested~\cite{Wood2004,Allmond2008,Allmond2009,Allmond2010}, which
provides new insights into the physics of triaxial rotations. As the
triaxial rotation is explicitly defined in the quadrupole-deformed
rotor, it is necessary to clarify the differences between the
triaxial dynamics generated by the different type of
quadrupole-deformed rotor Hamiltonian, especially by seeing that
both the rigid and irrotational type rotor are used to describe
nuclear collectivity~\cite{Zhang2014,Chen2014}. In this work, we
will present a systematical analysis of the similarities and
differences of level patterns and E2 transitional characters of the
irrotational type model and those of the rigid type model.

\begin{center}
\textbf{II Quadrupole-deformed ellipsoids and their moments of
inertia}
\end{center}\vskip.2cm

If only quadrupole-deformation is considered, the nuclear
 surface in the body-fixed frame (the principal axis system)
 may be described as~\cite{Bohrbook,Greinerbook}.
 \begin{equation}
 R(\theta,\phi)=R_0[1+\sum_\nu a_\nu Y_{2\nu}(\theta,\phi)]\, ,
 \end{equation}
 where $R_0$ is the radius of the nucleus with spherical shape, $\{a_\nu\}$ represent
 components of the quadrupole deformation with
 \begin{equation}
 a_1=a_{-1}=0,~~~~a_2=a_{-2}\, ,
 \end{equation}
 and $Y_{2\nu}(\theta,\phi)$ is the spherical harmonics.
 It is more convenient to use another set of parameters~\cite{Bohrbook,Greinerbook}
 introduced by A. Bohr defined via
 \begin{eqnarray}\nonumber
 &&a_0=\beta \mathrm{cos}\gamma\, ,\\
 &&a_2=a_{-2}=\frac{1}{\sqrt{2}}\beta \mathrm{sin}\gamma\, ,
 \end{eqnarray}
 where $\beta$ represents the total deformation with
 \begin{equation}
 \sum_\nu |a_\nu|^2=\beta^2\, ,
 \end{equation} and $\gamma$ represents the degree of
 triaxiality.

 The deviation of $R(\theta, \phi)$ from $R_{0}$ is
 given by
 \begin{eqnarray}
 &&\Delta R(\theta, \phi)=R(\theta, \phi)-R_0\, \\\nonumber
 &&=\sqrt{\frac{5}{16\pi}}R_0\beta[\mathrm{cos}\gamma(\mathrm{3cos}^2\theta-1)+\sqrt{3}\mathrm{sin}\gamma \mathrm{sin}^2\theta
 \mathrm{cos2}\phi]\, .
 \end{eqnarray}
 It can be proven that all quadrupole-deformed shapes can be
 covered by $\gamma$ within $[0, \frac{\pi}{3}]$.
 Thus the deviations of $R(\theta, \phi)$ from $R_{0}$
 along the principle axes
 \begin{eqnarray}\label{R}\nonumber
 &&\Delta R_1=R_1-R_0=R(\frac{\pi}{2},0)-R_0\, ,\\
 &&\Delta R_2=R_2-R_0=R(\frac{\pi}{2},\frac{\pi}{2})-R_0\, ,\\\nonumber
 &&\Delta R_3=R_3-R_0=R(0,\phi)-R_0\,
 \end{eqnarray}
 can be summarized as
 \begin{equation}
 \Delta R_\lambda=\sqrt{\frac{5}{4\pi}}R_0 \beta \mathrm{cos}
 (\gamma-\frac{2\lambda\pi}{3})~~{\rm with}~~\lambda=1,~2,~3\, .
 \end{equation}
 Specifically, one may find \begin{eqnarray}
 \Delta R_1=\Delta R_2=-\sqrt{\frac{5}{16\pi}}R_0\beta\,,~~\Delta R_3=\sqrt{\frac{5}{4\pi}}R_0\beta\,
 \end{eqnarray}
 at $\gamma=0$;
 \begin{eqnarray}
 \Delta R_1=\Delta R_3=\sqrt{\frac{5}{16\pi}}R_0\beta\,, ~~\Delta R_2=-\sqrt{\frac{5}{4\pi}}R_0\beta\,
 \end{eqnarray}
 at $\gamma=\frac{\pi}{3}$;
 \begin{eqnarray}
 \Delta R_1=0\, ,~~\Delta R_2=-\Delta R_3=-\sqrt{\frac{15}{16\pi}}R_0\beta\,
 \end{eqnarray}
 at $\gamma=\frac{\pi}{6}$.
 If only $\beta>0$ is allowed, the above results indicate that $\gamma=0$ represents the prolate
shape, $\gamma=\frac{\pi}{3}$ represents the oblate shape, and
$\gamma=\frac{\pi}{6}$ corresponds to the most triaxial shape.

 Although the deformation parameters $\beta$ and $\gamma$ are
 not observables, one can judge the geometrical shape of a
 deformed nucleus from its rotational spectrum if and only if the nucleus is assumed to be rigid.
 The rotor Hamiltonian is given by~\cite{Davydov1958,Wood2004}
 \begin{equation}\label{Hr}
 H_{\mathrm{rot}}=\frac{1}{2\Im_1}L_1^2+\frac{1}{2\Im_2}L_2^2+\frac{1}{2\Im_3}L_3^2\, ,
 \end{equation}
 where $L_\alpha$ is the projection of the angular momentum along the
 $\alpha$-th body-fixed principal axis and $\Im_\alpha$ is the
 corresponding moment of inertia. In the following,
 only rigid or irrotational ellipsoid is assumed
 to discuss the $\beta$- and $\gamma$-dependence of the moments of inertia.

 For a rigid ellipsoid with uniform mass density distribution, the moments of inertia
 along the principle axes may be expressed as
 \begin{eqnarray}\label{Ga}\nonumber
&&\Im_1=\Gamma_1=\frac{M}{5}({R_2^2+R_3^2})\,,~~\Im_2=\Gamma_2=\frac{M}{5}({R_1^2+R_3^2})\, ,\\
&&\Im_3=\Gamma_3=\frac{M}{5}({R_1^2+R_2^2})\, ,
 \end{eqnarray}
 where $M$ is the mass of the ellipsoid.
 Substituting $R_i$ with $i=1,~2,~3$  given by Eq.~(\ref{R}) into (\ref{Ga}),
 one has
 \begin{eqnarray}\nonumber
 &&\Gamma_1=2C[1+D\mathrm{cos}(\gamma+\frac{\pi}{3})+D^2(\frac{1}{4}\mathrm{cos}(2\gamma-\frac{\pi}{3})+\frac{1}{2})]\, ,\\\nonumber
 &&\Gamma_2=2C[1+D\mathrm{cos}(\gamma-\frac{\pi}{3})+D^2(\frac{1}{4}\mathrm{cos}(2\gamma+\frac{\pi}{3})+\frac{1}{2})]\, ,\\\nonumber
 &&\Gamma_3=2C[1+D\mathrm{cos}(\gamma-\pi)+D^2(\frac{1}{4}\mathrm{cos}(2\gamma+\pi)+\frac{1}{2})]\,, \\
 \end{eqnarray}
 where $C=\frac{MR_0^2}{5}$ and $D=\sqrt{\frac{5\beta^2}{4\pi}}$, which can be further simplified as
 \begin{equation}\label{Trigid}
 \Gamma_\lambda=2C[1-D\mathrm{cos}(\gamma-\frac{2\lambda\pi}{3})-\frac{D^2}{2}\mathrm{cos^2}(\gamma+\frac{\lambda\pi}{3})+\frac{3D^2}{4}]\,
 \end{equation}
 with $\lambda=1,~2,~3$.

 It can easily be found that
 $\Gamma_1=\Gamma_2>\Gamma_3$ at $\gamma=0$ corresponding to
 the prolate shape, $\Gamma_1=\Gamma_3<\Gamma_2$ at $\gamma=\frac{\pi}{3}$ corresponding to
 the oblate shape, and $\Gamma_2>\Gamma_1>\Gamma_3$ at $\gamma=\frac{\pi}{6}$
 corresponding to the most triaxial shape.
 It is obvious that the dynamical shape characterized by the moments of inertia $\frac{1}{2\Im_\alpha}$
 with $\alpha=1$, $2$, $3$  is always
 consistent with the geometric shape
 characterized by the Bohr variable $\gamma$ for the rigid type ellipsoid.
 Moreover, when $\Gamma_1=\Gamma_2$ or $\Gamma_1=\Gamma_3$,
 the spectrum of (\ref{Hr}) obeys the rotational $L(L+1)$-law within each
 rotational band. Therefore, the spectrum of the prolate or the oblate rigid ellipsoid
 is called regular.

On the other hand, for an irrotational ellipsoid with the same mass
density distribution, one may write the moments of
 inertia along the principle axes as~\cite{Bohrbook}
 \begin{eqnarray}\nonumber
 &&\Im_{1}=\Gamma_1^\prime=\frac{M}{5}\frac{(R_2^2-R_3^2)^2}{R_2^2+R_3^2}\,,~
 \Im_{2}=\Gamma_2^\prime=\frac{M}{5}\frac{(R_1^2-R_3^2)^2}{R_1^2+R_3^2}\, ,\\
 &&\Im_{3}=\Gamma_3^\prime=\frac{M}{5}\frac{(R_1^2-R_2^2)^2}{R_1^2+R_2^2}\, .
 \end{eqnarray}
 Specifically, the moments of inertia of the irrotational ellipsoid
 shown in (15) may be expressed as functions of $\beta$ and $\gamma$
 according to  Eq.~(\ref{R}) as
 \begin{eqnarray}\nonumber
 &&\Gamma_1^\prime=\frac{C[2\sqrt{3}D\mathrm{sin}(\gamma-\frac{2\pi}{3})-
 \frac{\sqrt{3}D^2}{2}\mathrm{sin}(2\gamma+\frac{2\pi}{3})]^2}{2+D^2-2D\mathrm{cos}(\gamma-\frac{2\pi}{3})-\frac{D^2}{2}\mathrm{cos}(2\gamma+\frac{2\pi}{3})}\, ,\\\nonumber
 &&\Gamma_2^\prime=\frac{C[2\sqrt{3}D\mathrm{sin}(\gamma-\frac{4\pi}{3})-
 \frac{\sqrt{3}D^2}{2}\mathrm{sin}(2\gamma+\frac{4\pi}{3})]^2}{2+D^2-2D\mathrm{cos}(\gamma-\frac{4\pi}{3})-\frac{D^2}{2}\mathrm{cos}(2\gamma+\frac{4\pi}{3})}\, ,\\\nonumber
 &&\Gamma_3^\prime=\frac{C[2\sqrt{3}D\mathrm{sin}\gamma-
 \frac{\sqrt{3}D^2}{2}\mathrm{sin}2\gamma]^2}{2+D^2-2D\mathrm{cos}\gamma-\frac{D^2}{2}\mathrm{cos}2\gamma}\, ,\\
 \end{eqnarray}
 which may be  rewritten uniformly as
 \begin{equation}\label{Tirro0}
 \Gamma_\lambda^\prime=\frac{C[2\sqrt{3}D\mathrm{sin}(\gamma-\frac{2\lambda\pi}{3})-
 \frac{\sqrt{3}D^2}{2}\mathrm{sin}(2\gamma+\frac{2\lambda\pi}{3})]^2}
 {2+D^2-2D\mathrm{cos}(\gamma-\frac{2\lambda\pi}{3})-\frac{D^2}{2}\mathrm{cos}(2\gamma+\frac{2\lambda\pi}{3})}\,
 \end{equation}
 for $\lambda=1,~2,~3$.
 Since $D$ or $\beta$ is usually a small quantity, to the leading order of $D$,
 the moments of  inertia  of the irrotational ellipsoid are given by
 \begin{equation}\label{Tirro-1}
 \Gamma_\lambda^\prime=6CD^2\mathrm{sin}^2(\gamma-\frac{2\lambda\pi}{3})\, .
 \end{equation}
 By submitting the collective mass parameter defined as
 $B=\frac{3}{8\pi}MR_0^2$, one may get the familiar
 form  with~\cite{Greinerbook}
 \begin{equation}\label{Tirro}
 \Gamma_\lambda^\prime=4B\beta^2\mathrm{sin}^2(\gamma-\frac{2\lambda\pi}{3})\, ,
 \end{equation}
 which can also be obtained from the derivation shown in ~\cite{Bohrbook,Greinerbook} by using
 the quantization procedure.

 According to (\ref{Tirro}), in comparison to the rigid type shown in (14),
 $\Gamma_1^\prime=\Gamma_2^\prime=3B\beta^2$ and $\Gamma_3^\prime=0$  in
 the prolate case at $\gamma=0$, $\Gamma_1^\prime=\Gamma_3^\prime=3B\beta^2$ and $\Gamma_2^\prime=0$  in
 the oblate case at $\gamma=\frac{\pi}{3}$, and $\Gamma_2^\prime=\Gamma_3^\prime=B\beta^2$ and $\Gamma_1^\prime=4B\beta^2$
 in the most triaxial case at $\gamma=\frac{\pi}{6}$. It should be noted that the moments of inertia of
 the irrotational type ellipsoid at $\gamma=\pi/6$ is symmetric with respect to the 2nd and 3rd principal axes exchange
 though the geometric shape is most triaxial according to (10).
It is clear that the dynamical shape characterized by the moments of
inertia $\frac{1}{2\Im_\alpha}$ is inconsistent with the geometric
shape characterized by the Bohr variable $\gamma$ for  the
irrotational type ellipsoid in either the oblate case or the most
triaxial case.

\begin{center}
\vskip.2cm\textbf{III Comparison of the rigid and irrotational
ellipsoid dynamics}
\end{center}\vskip.2cm

The quantum dynamics of a rotor described by (\ref{Hr}) is
 determined by relative magnitudes of the moments of inertia.
 As a consequence, differences and similarities in the spectral patterns and E2 transitional characters
 of the rigid ellipsoid and those of the irrotational ellipsoid
 can be analyzed accordingly.

 It should be noted that, no matter whether a quantum ellipsoid with
 exact axial-symmetry is rigid or irrotational, its arbitrary rotation
 around its axial-symmetry axis is quantum mechanically undetectable
 due to the additional $O(2)$ symmetry.
 In this extreme case, its spectrum involves only $K=0$ band as clarified in ~\cite{Greinerbook}, of which
 the levels obey the $L(L+1)$-law, where $L$ and $K$ is the total angular momentum and its projection onto the symmetric principal
 axis.

 As shown in the previous section, the axially-symmetric situations
 occur at $\gamma=0$ and $\pi/3$ corresponding to the prolate and
 oblate shape, respectively, for the rigid type and at $\gamma=0$,
 $\pi/6$, and $\pi/3$ for the irrotational type. As a result, one can
 not tell whether the ellipsoid is rigid or irrotational from its
 spectrum when $\gamma=0$ or $\gamma=\pi/3$. Although axial-symmetric
 situation is unrealistic in describing rotational motion of deformed
 nuclei, a comparison of spectral characters of the rigid ellipsoid
 to the irrotational one in this extreme case is instructive.
 Actually, up to a scaling factor, spectra of the two types of
 ellipsoid are the same at $\gamma=0$ because the relation
$\Im_1=\Im_2>\Im_3$ is satisfied for both types.
 Moreover, there is no distinction of  the irrotatioal ellipsoid at $\gamma=0$
 from that at $\gamma=\pi/3$ in spectra because energy levels generated
 from the two are the same.
 In contrast, the scaling of excitation energies of the prolate ellipsoid
 is different from that of the oblate one in the rigid case as shown
 from the moments of inertia given in (14).
 An interesting point is that, up to a scaling factor,
 spectrum of the irrotational ellipsoid  in the most triaxial
 case at $\gamma=\pi/6$ coincides with that of the rigid one
 in the oblate case at $\gamma=\pi/3$.
 Therefore, spectral characters of ellipsoids of different type but of different
 geometric shape may be quite similar, even when the axial-symmetry is slightly broken.

 In fact, if only the leading order of $D$ is considered,
 the moments of inertia for the rigid case shown in (\ref{Trigid}) may be
 expressed as
 \begin{equation}\label{TrigidII}
 \Gamma_\lambda=2C[1-D\mathrm{cos}(\gamma-\frac{2\lambda\pi}{3})]\,
 \end{equation}
 since $D=\sqrt{\frac{5\beta^2}{4\pi}}$ is generally a small
 quantity with $0<D<1$,
 while the moments of inertia for irrotational case shown
 in (\ref{Tirro-1}) can be rewritten as
 \begin{equation}\label{Tirro-2}
 \Gamma_\lambda^\prime=3CD^2[1-\mathrm{cos}(2\gamma-\frac{4\lambda\pi}{3})]\, .
 \end{equation}
 In comparison of (\ref{TrigidII}) with (\ref{Tirro-2}), it is obvious
 that, up to a scaling factor, there is the one-to-one correspondence
 between the moments of inertia of rigid type shown in (\ref{TrigidII}) at
 $\gamma=2t$ with $0\leq t\leq\pi/6$
 and those of irrotational type shown in (\ref{Tirro-2})
 with the 1st and the 2nd principal axis exchange at $\gamma=t$
 when $D=1$.

 \begin{figure*}
\begin{center}
\resizebox{0.8\textwidth}{!}{
\includegraphics{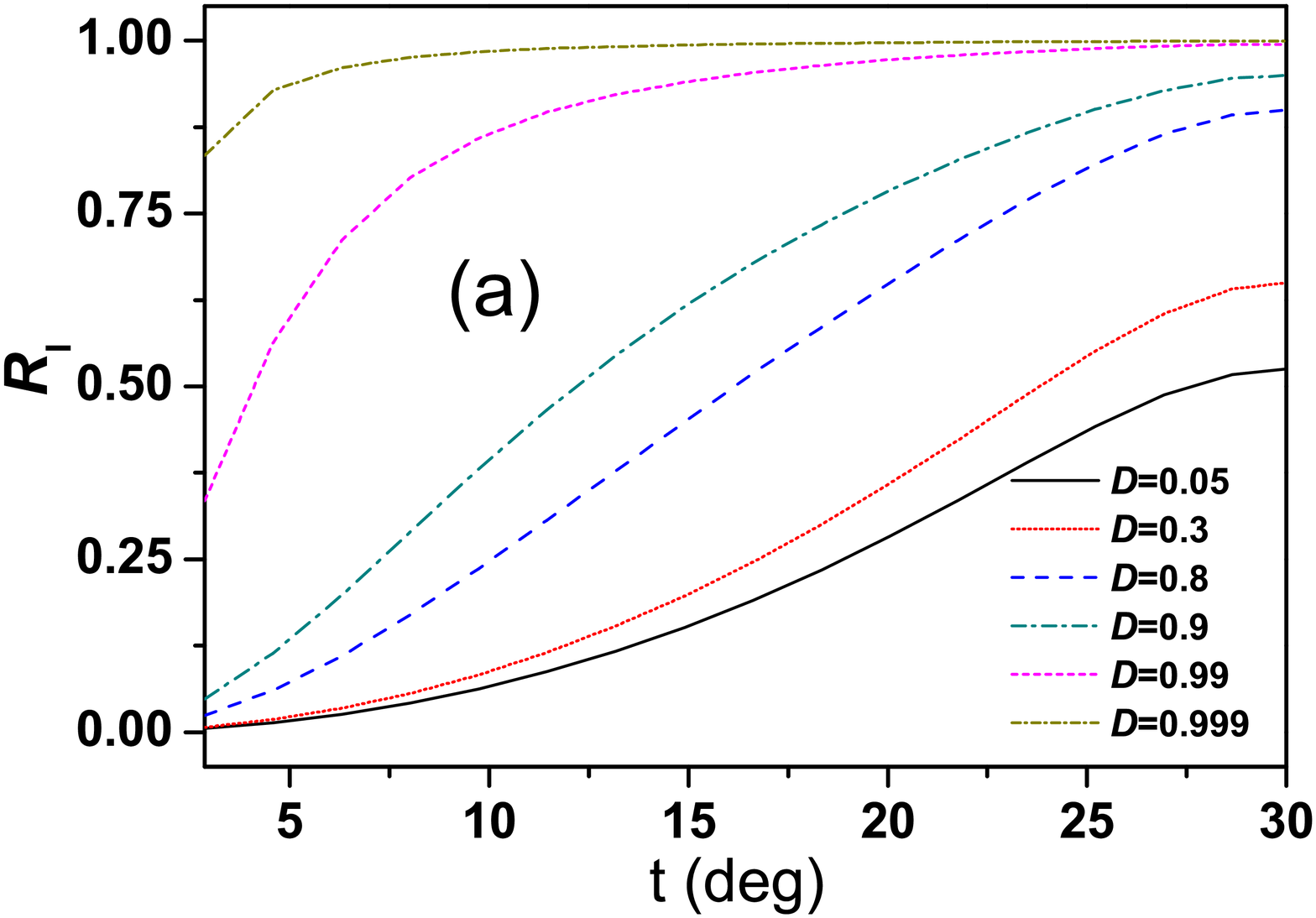}
\includegraphics{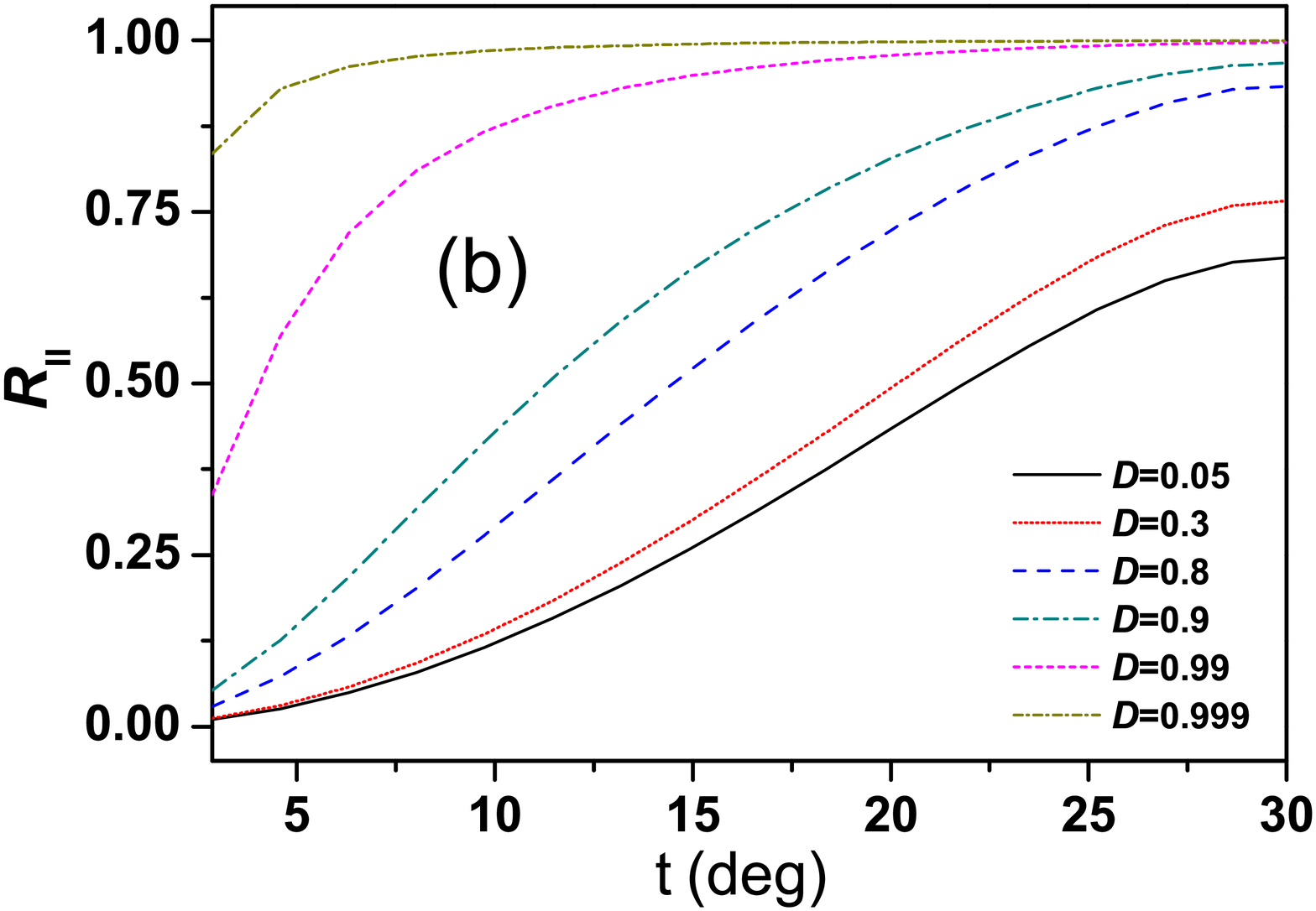}}
\caption{The quantities $R_\mathrm{I}$ and $R_\mathrm{II}$ with
$D=0.05,~0.3,~0.8,~0.9~,0.99,~0.999$ are shown as functions of $t$
(in degree).} \label{F0}
\end{center}
\end{figure*}

 Furthermore, it can be verified that such a correspondence is preserved even for $0<D<1$.
 For example, it can be shown that relative magnitudes of the moments of inertia given by (\ref{TrigidII}) at
 $\gamma=\pi/6$ are similar to those shown in (\ref{Tirro-2}) at $\gamma=\pi/12$
 since the relations $\Im_a>\Im_b>\Im_c$ and
 $(\Im_a+\Im_c)/2=\Im_b$ are satisfied for both cases with any given $D$ value,
 where $(a,~b,~c)$ represents $(1,~2,~3)$ for the rigid case and
 $(2,~1,~3)$ for the irrotational case.
 To illustrate the
 effect of $D$ (or $\beta$) on the level structure, we take the two
 typical energy ratios, $E_{2_2^+}/E_{2_1^+}$ and
 $E_{3_1^+}/E_{2_1^+}$, as the examples.
 Explicitly, the two energy ratios can be analytically expressed as
 \begin{eqnarray}\label{Riiro}
 &R_{\mathrm{irroI}}\equiv\frac{E_{2_2^+}}{E_{2_1^+}}=\frac{3+\sqrt{5+4\mathrm{cos}(6\gamma)}}{3-\sqrt{5+4\mathrm{cos}(6\gamma)}}\\
 &R_{\mathrm{irroII}}\equiv\frac{E_{3_1^+}}{E_{2_1^+}}=\frac{6}{3-\sqrt{5+4\mathrm{cos}(6\gamma)}}\,
 \end{eqnarray}
 for the ones solved from the irrotational type rotor and
 \begin{eqnarray}\label{Rrig}
 &R_{\mathrm{rigI}}\equiv\frac{E_{2_2^+}}{E_{2_1^+}}=\frac{4-D^2+\sqrt{D^2(4+D^2+4D\mathrm{cos}(3\gamma))}}{4-D^2-\sqrt{D^2(4+D^2+4D\mathrm{cos}(3\gamma))}}\\
 &R_{\mathrm{rigII}}\equiv\frac{E_{3_1^+}}{E_{2_1^+}}=\frac{2(4-D^2)}{4-D^2-\sqrt{D^2(4+D^2+4D\mathrm{cos}(3\gamma))}}\,
 \end{eqnarray}
 for those solved from the rigid type rotor. It is clear that the energy ratios
 in the irrotational case depend on only the $\gamma$ variable but the
 ratios in the rigid case depend on both the $\gamma$ and
 $\beta$ variables. Particularly, it can be fount that the ratios in the rigid case with $\gamma=2t$ may equal to those in the rigid case with $\gamma=t$
 in the $D=1$ limit. Further, one can define the quantities, $R_\mathrm{I}$ and $R_\mathrm{II}$, as $R_\mathrm{I}(t)=R_{\mathrm{rigI}}(\gamma=2t)/R_{\mathrm{irroI}}(\gamma=t)$ and
 $R_\mathrm{II}(t)=R_{\mathrm{rigII}}(\gamma=2t)/R_{\mathrm{irroII}}(\gamma=t)$ to test the $D$ dependence of the difference
 in between the two types of rotor, and the calculated results are given in Fig.~\ref{F0}. As clearly seen from Fig.~\ref{F0}, the values of both $R_\mathrm{I}$ and
 $R_\mathrm{II}$ increase monotonically as increasing of $t$ and $D$,
 which indicates that the difference in the energy ratios between
 the two types model defined above may become very small for large $\beta$ and $\gamma$ deformations.

\begin{figure}
\begin{center}
\resizebox{0.5\textwidth}{!}{
\includegraphics{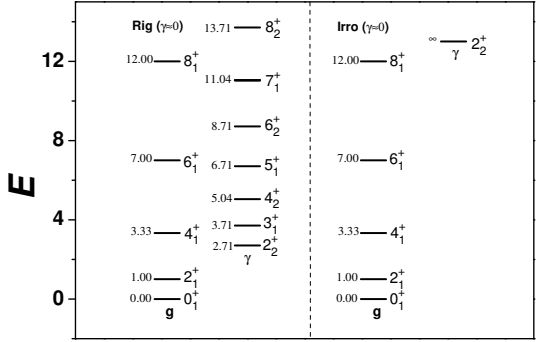}}
\caption{Some low-lying levels in the ground- and $\gamma$-band of
the prolate case for the rigid (Rig) ellipsoid (left) and the
irrotational (Irro.) one (right), in which $L_\xi^+$ denotes the
$\xi$-th positive-parity state with the angular momentum quantum
number $L$, and all the energy values have been normalized to the
first $2^+$ energy in the ground-band.} \label{F1}
 \end{center}
 \end{figure}

A further comparison between the two types of ellipsoid should be
made for both level patterns and E2 transitional characters.
 To obtain the energy levels and E2 transitional rates,
 numerical diagonalization of  the Hamiltonian (\ref{Hr}) should be carried out.
 Eigenfunctions  $\Psi_{LM}^K$ of the general rotor  Hamiltonian (\ref{Hr})
 may be expanded in terms of the Wigner $D$-functions with
  \begin{equation}
 \Psi_{LM}^\xi=\sum_KC_K^\xi\Psi_{LM}^K\, ,
 \end{equation}
 where $M$ is the quantum number of the angular momentum projection onto the third axis
 in the laboratory frame, $\{C_K^\xi\}$ are the expansion coefficients, $\xi$ is an
 additional quantum number needed to label different eigenstates
 with the same quantum number $L$ and $M$, and
 \begin{eqnarray}\label{basis}\nonumber
 \Psi_{LM}^{K}=&\sqrt{\frac{2L+1}{16\pi^2(1+\delta_{K0})}}[D_{M,K}^{(L)*}(\theta_1,\theta_2,\theta_3)\\
 &+(-1)^{L}D_{M,-K}^{(L)*}(\theta_1,\theta_2,\theta_3)]\,,
 \end{eqnarray}
 in which $D_{M,K}^{(L)}(\theta_1,\theta_2,\theta_3)$ is the Wigner D-function of Euler angles $\theta_1$, $\theta_2$, and $\theta_3$.
 The Hamiltonian (\ref{Hr}) under the basis spanned by (\ref{basis}) are block-diagonalized.
 The block-diagonalized result is due to the invariance of (\ref{Hr}) under rotations by $\pi$ around the principal
 axes~\cite{CDL1988}. These rotations, which can be written as $T_\alpha=e^{-i\pi L_\alpha}$ with $\alpha=1$,~$2$,~ and $3$,
 together with the identity operation, generate the Vieregruppe ($D_2$)
 group. The invariance means that $[H_{\mathrm{rot}},~T_\alpha]=0$ for $\alpha=1$,~$2$,~ and $3$.
 Generally, Wavefunctions that carry the irreps of the $D_2$ can be
 constructed by a combination of Wigner $D$-functions with
 \begin{eqnarray}\label{basisII}\nonumber
 \hskip -1cm\Psi_{LM}^{\lambda\mu K}=&\sqrt{\frac{2L+1}{16\pi^2(1+\delta_{K0})}}[D_{M,K}^{(L)*}(\theta_1,\theta_2,\theta_3)\\
 &+(-1)^{\lambda+\mu+L}D_{M,-K}^{(L)*}(\theta_1,\theta_2,\theta_3)]\,
 ,
 \end{eqnarray}
 where $\lambda$ and $\mu$ are integers. The  $D_2$ group has four
 representations denoted as $A$, $B_1$,~$B_2$, and $B_3$, respectively, in which
only the $A$-type representation is allowed for even-even
nuclei~\cite{CDL1988}.
 In the $A$-type case, both $\lambda$ and $\mu$ should be
 taken as even integers, by which (\ref{basisII}) is reduced to (\ref{basis})
 with $K=0$ or $K=\mathrm{even}$, in which only positive $K$ values need to be
 considered.
 The multiplicity of $L$ is
 given as $(L+2)/2$ for $L=\mathrm{even}$ and $(L-1)/2$ for $L=\mathrm{odd}$.
 As shown in~\cite{Leschber1987}, the allowed $L$ and $K$ are  $L=0,~2,~4,~6,\cdots$, for $K=0$ and
 $L=K,~K+1,~K+2,~K+3,\cdots$, for $K=\mathrm{even}$.

\begin{table}[htb]
\caption{Some typical $B(E2)$ values  for the two types of ellipsoid
in the prolate case corresponding to the case shown in
Fig.~\ref{F1}, where all transitions are normalized to $B(E2;
2_g\rightarrow0_g)$, of which $L_g$ and $L_\gamma$ denote the states
with angular momentum quantum number $L$ in the ground-band and
those in the $\gamma$-band, respectively. In the calculations, the
$\gamma$ value in the quadrupole operator (\ref{Q}) has been taken
the same as that used in the corresponding moments of inertia.}
\begin{center}
\label{T1}
\begin{tabular}{cccccc}\hline\hline
$L_i\rightarrow L_f$ & Rig & Irro&$L_i\rightarrow L_f$ & Rig & Irro
\\\hline
$2_g\rightarrow0_g$&100&100&$2_\gamma\rightarrow0_g$&0&-\\
$4_g\rightarrow2_g$&143&143&$2_\gamma\rightarrow2_g$&0&-\\
$6_g\rightarrow4_g$&157&157&$3_\gamma\rightarrow2_\gamma$&179&-\\
$8_g\rightarrow6_g$&165&165&$4_\gamma\rightarrow3_\gamma$&133&-\\
\hline\hline
\end{tabular}
\end{center}
\end{table}

The $D_2$ symmetry holds for both the asymmetric and the symmetric
cases of  (\ref{Hr}). In the dynamically axially-symmetric cases,
however, in addition to the $D_2$ symmetry, there is the additional
$O(2)$ symmetry. The $O(2)$ group consists of rotations around the
symmetric principal axis. When $\Im_{1}=\Im_2\neq\Im_{3}$ for
example, the eigenfunctions of (\ref{Hr}) in this case is those
shown in (\ref{basis}). However, (\ref{Hr}) in this case should also
be invariant under arbitrary rotation round the 3rd principal axis,
namely, $[H_{\mathrm{rot}},~e^{-i\phi L_3}]=0$ for arbitrary
$\phi\in[0,2\pi]$. The additional $O(2)$ symmetry requires that only
$K=0$ is allowed, because the eigenfunctions given by (\ref{basis})
under the $O(2)$ rotation transforms as
\begin{eqnarray}\label{basisIII}\nonumber
e^{-i\phi
L_3}\Psi_{L,M}^K=&\sqrt{\frac{2L+1}{16\pi^2(1+\delta_{K0})}}[e^{-i\phi
K} D_{M,K}^{*(L)}(\theta_1,\theta_2,\theta_3)\\
&+(-1)^{L}e^{i\phi K}D_{M,-K}^{*(L)}(\theta_1,\theta_2,\theta_3)]\,
,
\end{eqnarray}
which is invariant only when $K=0$. The additional $O(2)$ symmetry
answers
 why only the ground-band with $K=0$ emerges for the axially-symmetric rotor Hamiltonian~\cite{Greinerbook}.
Actually, the axially-symmetric rotor Hamiltonian in this case is
invariant under the $O(2)\overline{\bigotimes} D_2$ transformation,
where $\overline{\bigotimes}$ stands for the semi-direct product.

In the following, we closely compare the rotational features of
ellipsoid of the rigid type with that of the
 irrotational type at several special $\gamma$ points, of which situations with exact axial-symmetry
 is avoid because the exact axial-symmetry is unrealistic in describing
 deformed nuclei. In order to avoid the exact axial-symmetry,
 a very small quantity $\varepsilon$ is always assumed to be added
 to the $\gamma$ values corresponding to the exact axially-symmetric cases,
 though the results calculated with $\gamma+\varepsilon$ and
 those with $\gamma$ are approximately taken as the same. Moreover,
 besides the ground-band, other bands in this case also show up
 with the near axial-symmetry assumption.

 \begin{figure}[htb]
\begin{center}
\resizebox{0.5\textwidth}{!}{
\includegraphics{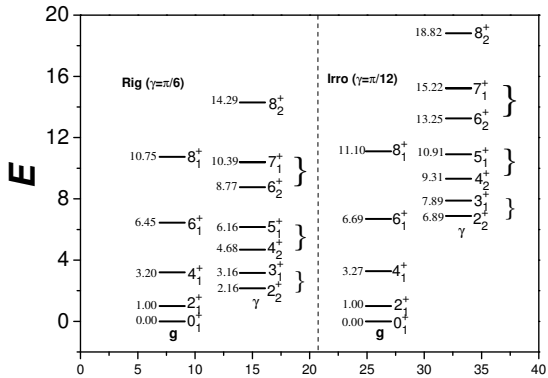}}
\caption{The same as Fig.~\ref{F1} but for the two types of
ellipsoid in the triaxial case.} \label{F2}
\end{center}
\end{figure}

 Some low-lying levels in both the ground-band and the $\gamma$-band
 of the rigid ellipsoid at $\gamma=2t$ and those of the irrotational
 ellipsoid at $\gamma=t$  with $t\approx0$, $t=\pi/12$, and $t\approx\pi/6$ are shown in
 Figs. 1-3, respectively, where $\beta=1.0$ is set for all these cases.
 B(E2) values of the above cases for both the intra- and inter-band transitions
 are calculated according to
 \begin{equation}
 B(E2;L_i\rightarrow L_f)=\frac{|\langle L_f\|\hat{Q}\|L_i\rangle|^2}{2L_i+1}\, ,
 \end{equation}
 where the quadrupole operator is given as~\cite{Bohrbook,Greinerbook}
 \begin{equation}\label{Q}
 \hat{Q}_u={3Ze\over{4\pi}}R_{0}^{2}~
 \beta[\mathrm{cos}(\gamma)D_{u,0}^{(2)}+\frac{1}{\sqrt{2}}\mathrm{sin}(\gamma)(D_{u,2}^{(2)}+D_{u,-2}^{(2)})]\,
 ,
 \end{equation}
 in which $e$ is assumed to be the effective charge.
Unless specified separately, the $\beta$ and $\gamma$ values in
(\ref{Q}) are taken the same as those in the moments of inertia for
a given type of ellipsoid. Some typical B(E2) values of both the
intra- and inter-band transitions of the rigid ellipsoid at
$\gamma=2t$ and those of the irrotational ellipsoid at $\gamma=t$
for $t\approx0$, $t=\pi/12$, and $t\approx\pi/6$ are shown in Tables
1-3, respectively.

\begin{table}[htb]
\caption{The same as  Table.~\ref{T1} but for the triaxial case
corresponding to Fig.~\ref{F2}.}
\begin{center}
\label{T2}
\begin{tabular}{cccccc}\hline\hline
$L_i\rightarrow L_f$ & Rig & Irro &$L_i\rightarrow L_f$ & Rig
&Irro\\\hline
$2_g\rightarrow0_g$&100&100&$2_\gamma\rightarrow0_g$&54&6\\
$4_g\rightarrow2_g$&109&145&$2_\gamma\rightarrow2_g$&20&15\\
$6_g\rightarrow4_g$&99&165&$3_\gamma\rightarrow2_\gamma$&179&179\\
$8_g\rightarrow6_g$&97&179&$4_\gamma\rightarrow3_\gamma$&188&123\\
\hline\hline
\end{tabular}
\end{center}
\end{table}

As shown in Fig.~\ref{F1}, both the ground-band
 and the $\gamma$-band for the two types of
 ellipsoids of the prolate shape follow the $L(L+1)$-law
 exactly except that the band head energy of the $\gamma$-band is
 infinite in the irrotational case due to
 $\Gamma_3^\prime=0$ at $\gamma\approx0$.
 Notably, $E_{4_g}/E_{2_g}=3.33$ is the
 direct evidence that the levels in the ground-band obey the $L(L+1)$-law.
 For the $\gamma$-band, it is convenient to use the ratio defined as
 $R_\gamma=\frac{E_{4_\gamma}-E_{2_\gamma}}{E_{3_\gamma}-E_{2_\gamma}}$,
 of which the ratio $R_\gamma=2.33$ if the $L(L+1)$-law is satisfied.
 It should be
 noted that levels in the ground-band and those in the $\gamma$-band in the prolate case
 are grouped by those with
 $K^\pi=0^+$ and those with $K^\pi=2^+$, respectively,
 where $K$ is the projection of angular momentum onto
 the 3rd principle axis.
 In this case, the Hamiltonian is nearly axially-symmetric~\cite{Bohrbook}
 because $\Im_1\approx\Im_2>\Im_3$ is always satisfied for both the rigid and irrotational
 type when $\gamma\approx0$.
 Furthermore, the normalized results given in Table~\ref{T1}
 show that the allowed transitional rates for the two types of ellipsoid in the prolate case
 are also the same, while the inter-band transitions for the irrotational type are prohibited.

\begin{figure}[htb]
\begin{center}
\resizebox{0.5\textwidth}{!}{
\includegraphics{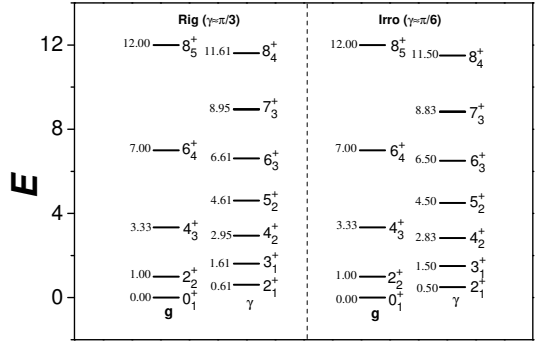}}
\caption{The same as Fig.~\ref{F1} but for the two types of
ellipsoid at $t\approx\pi/6$.} \label{F3}
\end{center}
\end{figure}

For the triaxial case, the results calculated
 for the rigid type at $\gamma=\pi/6$ and the
 corresponding results calculated for the irrotational type at $\gamma=\pi/12$ are given in
 Fig.~\ref{F2} and Table~\ref{T2}.
 It can be seen from Fig.~\ref{F2}
 that the level patterns for the rigid type is quite similar to those for the irrotational
 type except that the $\gamma$-band head energy is relatively high.
 Particularly, the odd-even staggering, which is regarded as a
 signature of a triaxial rotor~\cite{Smirnov2000,Zamfir1991},
 appears in the $\gamma$-band of both types of ellipsoid
 as indicated by braces in Fig.~\ref{F2}.
 In addition, $R_\gamma=2.52$
 for the rigid type and $R_\gamma=2.42$ for the irrotational type,
 which indicates that levels in the $\gamma$-band
 of the two types of ellipsoid noticeably deviate from the $L(L+1)$-law. On the
 other hand,  the E2 transitional rates
given in Table~\ref{T2} show that the B(E2, $L+2\rightarrow L$)
value of intra-band transition in the ground-band gradually
decreases with the increasing of $L$ for the rigid case, but
increases noticeably with the increasing of $L$ for the irrotational
case.

\begin{table}[htb]
\caption{The same as Table~\ref{T1} but for the case with
$t\approx\pi/6$ corresponding to Fig.~\ref{F3}, where the values
given in column {$\mathrm{Irro_a}$ and $\mathrm{Irro_b}$} are
obtained with {$\gamma=-\pi/2$} and $\gamma=5\pi/3$ in the
quadrupole operator (27), respectively. In addition, all the results
have been normalized to $B(E2;2_g\rightarrow0_g)$ except for those
in column Irro, which are normalized to
$B(E2;2_\gamma\rightarrow0_g)$.}
\begin{center}
\label{T3}
\begin{tabular}{cccccccccc}\hline\hline
$L_i\rightarrow L_f$
&Rig&Irro&$\mathrm{Irro_a}$&$\mathrm{Irro_b}$&$L_i\rightarrow
L_f$&Rig&Irro&$\mathrm{Irro_a}$&$\mathrm{Irro_b}$\\\hline
$2_g\rightarrow0_g$&100&0&100&100&$2_\gamma\rightarrow0_g$&0&100&33&0\\
$4_g\rightarrow2_g$&140&1&155&140&$2_g\rightarrow2_\gamma$&0&143&48&0\\
$6_g\rightarrow4_g$&155&4&184&155&$3_\gamma\rightarrow2_\gamma$&179&0&179&179\\
$8_g\rightarrow6_g$&163&6&202&163&$4_\gamma\rightarrow3_\gamma$&131&1&146&131\\
\hline\hline
\end{tabular}
\end{center}
\end{table}

 The results calculated for the rigid type at
 $\gamma\approx\pi/3$ and the corresponding results for the
 irrotational type at $\gamma\approx\pi/6$ are shown in Fig.~\ref{F3}
 and Table~\ref{T3}.
 It is clearly shown in Fig.~\ref{F3} that the levels in the ground-band of
 both the rigid and irrotational type ellipsoid
 follow the $L(L+1)$-law exactly. Since $R_\gamma=2.33$
 for both types, the levels in the $\gamma$-band of both types of ellipsoid
 also follow the $L(L+1)$-law exactly.
 A notable feature of both types is that the levels with $L$ even in the $\gamma$-band
 are all lower in energy than the corresponding ones in the ground-band as shown in
 Fig.~\ref{F3}.
 It should be emphasized that the Hamiltonian (\ref{Hr}) in this case is nearly axially-symmetric
 because $\Gamma_1\approx\Gamma_3<\Gamma_2$ for the rigid type and
 $\Gamma'_2\approx\Gamma'_3<\Gamma'_1$ for the irrotational type.
 As a consequence, the levels with $\eta=0$ and those with
 $\eta=2$ for the rigid case are taken to be in the ground-band and in the
 $\gamma$-band respectively, where $\eta$ is the
 projection of angular momentum onto the 2nd principle axis.
 For the irrotational type, the levels with $\alpha=0$ and those with $\alpha=2$
 are taken to be  in the ground-band and in the $\gamma$-band respectively,
 where $\alpha$ is the projection of angular momentum onto the 1st principle axis because
 the 1st principal axis is the symmetric axis in this case.
 Thus, it is easy to understand why the levels in each band shown in Fig.~\ref{F3}
 all follow the $L(L+1)$-law exactly.
 In contrast to the prolate case shown in Table~\ref{F1},
 in which B(E2) values of the intra-band transitions within the ground-band for
 both rigid and irrotational type are the same,
 E2 transitional characters of the rigid and irrotational type
 shown in the first two columns labeled as Rig and Irro in Table~\ref{T3} are completely different.
 It is clearly shown for the rigid type case that the inter-band transitions between the ground- and $\gamma$-band
 are completely prohibited, but the intra-band transitions are allowed and noticeable,
 while the situation in the irrotational type case is completely the inverse, in which
 the intra-band transitions in each band are nearly prohibited,
 but the inter-band transitions between two bands are allowed and noticeable.

\begin{figure}[htb]
\begin{center}
\resizebox{0.5\textwidth}{!}{
\includegraphics{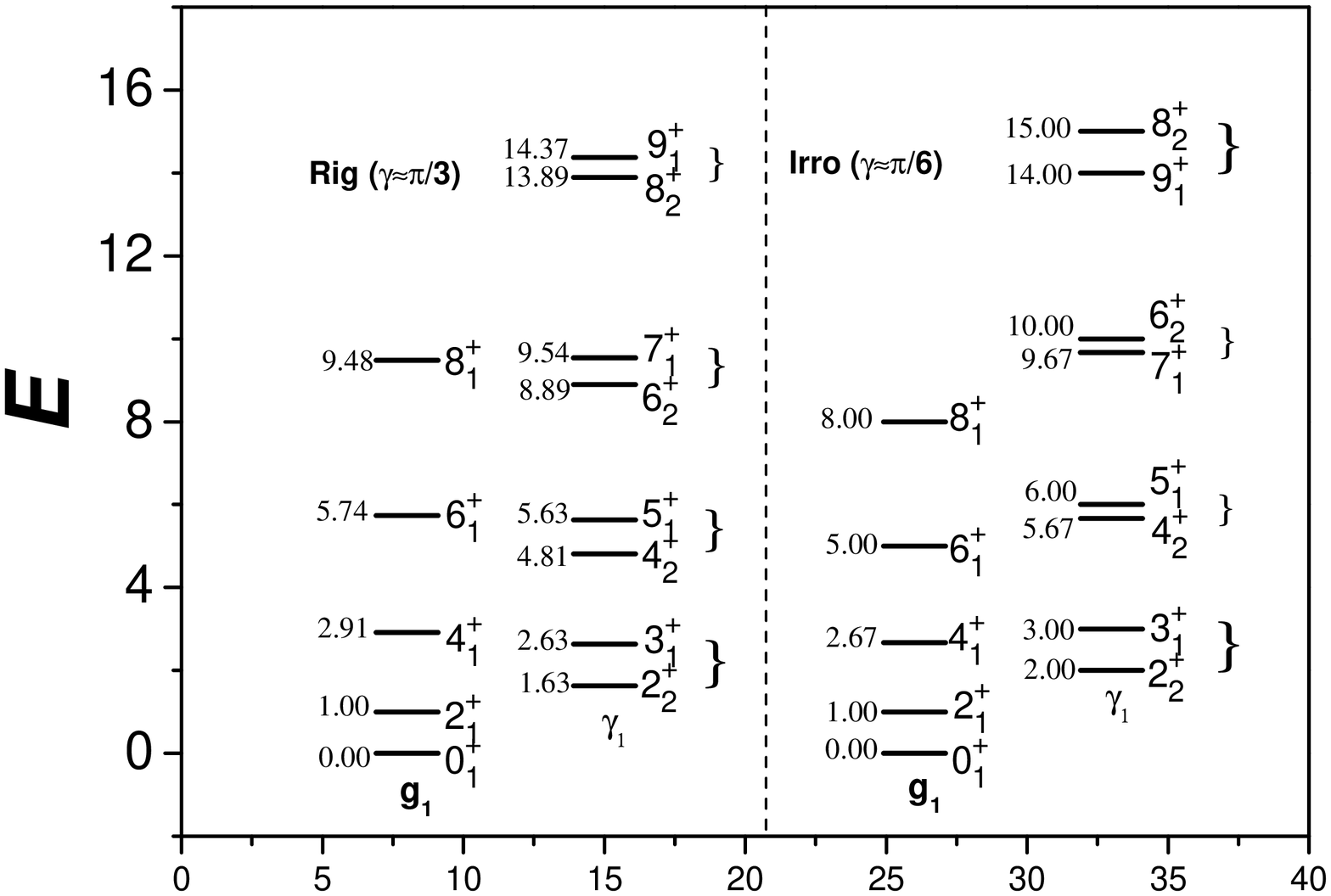}}
\caption{The same as Fig.~\ref{F3} but for the ground- and the
$\gamma$-band reassigned according to the energy value of the levels
with the same $L$ shown in the text, which are labeled by  $g_1$ and
$\gamma_1$ respectively. All the levels are normalized to the
$2_1^+$ energy in each case.} \label{F4}
\end{center}
\end{figure}

As mentioned previously, the $\gamma$ value in the electric
quadrupole operator has been taken the same as
 that in the moments of inertia, which assumes that the nuclear charge distribution follows its geometric shape exactly.
 If the electric quadrupole tensor and the inertia tensor can be independently parameterized as suggested in~\cite{Wood2004},
 one can then check what $\gamma$ value should be set in the electric quadrupole operator for the irrotational type to produce the
 the similar E2 transitional characters of the rigid type in the $t\approx\pi/6$ case.
 It can be verified that $\gamma=-\pi/2$ or $\gamma=5\pi/3$ may be adopted in the electric quadrupole operator (\ref{Q})
 for the irrotational type, of which $\gamma=t=\pi/6$ is taken for the moments of inertia in this irrotational type case.
 The corresponding B(E2) values are listed in the last two columns of Table~\ref{T3}.
 It should be noted that the electric quadrupole operator with {$\gamma=-\pi/2$} was also used in~\cite{Bonatsos2004}
 for this case.
 It is clearly shown in Table~\ref{T3} that the results
 obtained with these $\gamma$ values in the electric quadrupole operator
 for the irrotational type case are close to those of the rigid type case at $\gamma=2t=\pi/3$.
 Specifically, the results calculated with $\gamma=5\pi/3$
 in the electric quadrupole operator
 are even the same
 as the corresponding ones for the rigid type case.

\begin{table}[htb]
\caption{The same as Table~\ref{T3} but for transitions
corresponding to those shown in Fig.~\ref{F4}.}
\begin{center}
\label{T4}
\begin{tabular}{cccccccccc}\hline\hline
$L_i\rightarrow
L_f$&Rig&Irro&$\mathrm{Irro_a}$&$\mathrm{Irro_b}$&$L_i\rightarrow
L_f$ & Rig & Irro&$\mathrm{Irro_a}$&$\mathrm{Irro_b}$
\\\hline
$2_{g_1}\rightarrow0_{g_1}$&0&100&100&0&$2_{\gamma_1}\rightarrow0_{g_1}$&100&0&300&100\\
$4_{g_1}\rightarrow2_{g_1}$&0&141&165&0&$2_{\gamma_1}\rightarrow2_{g_1}$&0&143&143&0\\
$6_{g_1}\rightarrow4_{g_1}$&0&173&193&0&$3_{\gamma_1}\rightarrow2_{\gamma_1}$&0&179&536&0\\
$8_{g_1}\rightarrow6_{g_1}$&0&191&205&0&$4_{\gamma_1}\rightarrow3_{\gamma_1}$&131&1&438&131\\
\hline\hline
\end{tabular}
\end{center}
\end{table}

 Although the Hamiltonian for the irrotational type case at $\gamma=t=\pi/6$ is nearly axially-symmetric just
 as that for the rigid type case at {$\gamma\approx\pi/3$} corresponding to the oblate shape, the irrotational
 type at {$\gamma\approx\pi/6$}  is often
 referred to as being triaxial ~\cite{Bohrbook,Zamfir1991} because the corresponding
 geometrical shape is indeed most triaxial at {$\gamma=\pi/6$}.
 Actually, the levels shown in Fig.~\ref{F3} can also be
 regrouped into a new ground- and a new $\gamma$-band according to the
 energy value of the levels with the same $L$.
 Specifically, the new ground-band consists of the lowest levels with
 $L=0$ or even, while the new $\gamma$-band consist of the next to the lowest
 levels with $L={\rm even}$ or the lowest ones with $L={\rm odd}$, which
 is shown in Fig.~\ref{F4}.
 One can observe that the level pattern shown in Fig.~\ref{F3} and that shown in Fig.~\ref{F4}  are quite different.
 The odd-even staggering appears in the new $\gamma$-band for both
 types of ellipsoid, and the level ordering in the $\gamma$-band for the irrotational type case
 is even reversed, which are all considered to be signals of the triaxiality ~\cite{Smirnov2000,Zamfir1991}.
 For E2 transitions, it is shown in Table~\ref{T4}
 that the intra-band transitions in both the new ground- and the new $\gamma$-band
 are almost prohibited for the rigid type case, which applies
 to the irrotational type case shown in column $\mathrm{Irro_b}$ as well.
 In contrast, the intra-band transition rates in the ground-band shown in columns Irro and $\mathrm{Irro_a}$
 present the similar monotonic behavior, namely increasing with the increasing of $L$, as shown in
 Table~\ref{T4}.
 Hence, it is recognized that the
 triaxiality shown in Fig.~\ref{F4} emerges from the rearrangement of the levels
 of the oblate spectrum of the irrotational type case shown in Fig.~\ref{F3}, which
 results in a different band assignment.

\begin{center}
\vskip.2cm\textbf{V Summary}
\end{center}\vskip.2cm

 In summary, we have presented a detailed comparison of dynamical
shape characterized by the moments of inertia of the rigid type
ellipsoid to that characterized by those of the irrotational type.
It is shown that, up to an energy scaling factor, the level patterns
of the rigid ellipsoid at $\gamma=2t$ is similar to that of the
irrotational type at $\gamma=t$ to the leading order of the
deformation parameter. Numerical investigation on the excitaion
energies and B(E2) values for the two types of the model is also
carried out, in which the triaxial situations are particularly
emphasized. It is show that both level patterns and E2 transitional
characters of the two types of the model with prolate and triaxial
geometric shape are quite similar as shown in Fig.~\ref{F1},
Fig.~\ref{F2} and Table~\ref{T1}, Table~\ref{T2}. On the other hand,
it is shown in Fig.~\ref{F3} and Fig.~\ref{F4} that the level
patterns of the rigid type model with $\gamma=\pi/3$ is similar to
those of the irrotational type with $\gamma=\pi/6$.  However, if the
nuclear charge distribution is assumed to be different from its mass
distribution in the irrotational type model, it is shown in
Table~\ref{T3} and Table~\ref{T4} that a suitable choice of the
$\gamma$ value in the electric quadrupole operator for the
irrotational type model may result in similar E2 transitional rates
to those of the rigid type model, in which the  $\gamma$ value in
the moments of inertia is taken to be the same for both the
irrotational and the rigid type models. In addition, the results
also indicate that the excited levels in the rigid type model with
$\gamma=\pi/3$ and the irrotational type model with  $\gamma=\pi/6$
may be regrouped into the new ground- and $\gamma$-band, with which
the spectrum looks quite similar to that of a triaxial rotor. As a
result, similar rotational spectrum may be generated from different
type of the model with different $\gamma$ deformation parameter.
According to low-lying levels observed in most deformed nuclei, the
band assignment shown in Fig.~\ref{F4} seems more realistic than
that shown in Fig~\ref{F3}, namely the $\gamma$-band head energy
seems always higher than the energy of the first $2^+$ state in the
ground-band.

\begin{acknowledgments}
Support from the Natural Science Foundation of China (11375005,
11005056, 11175078).
\end{acknowledgments}


\begin{thebibliography}{99}

\bibitem{Kronig1937} R. L. Kronig and I. I. Rabi, Phy. Rev. {\bf 29}, (1927) 262.

\bibitem{Casimir1931} H. B. G. Casimir, {\it Rotationa of a rigid body in quantum mechanics} (Wolters, The Hague, 1931).

\bibitem{Bohr1952}A. Bohr, Konggl. Dan. Vid. Selsk. Mat.-fys. Medd. {\bf 26}, no. 14 (1952).

\bibitem{Davydov1958}A. S. Davydov, G. F. Filippov, Nucl.
                     Phys. \textbf{8}, (1958) 237.

\bibitem{Bohrbook} A. Bohr, B. R. Mottelson, {\it Nuclear Structure},
                     Vol. 1 and Vol. 2 (W. A. Benjamin, Inc., Reading, Ma 1975).

\bibitem{Greinerbook} W. Greiner, J. A. Maruhn, {\it Nuclear models} (Springer-Verlag, Berlin, 1996).


\bibitem{Ui1970}H. Ui, Prog. Theor. Phys. \textbf{44}, (1970) 153.

\bibitem{Leschber1987}Y. Leschber, J. P. Draayer, Phys. Lett. B \textbf{190}, (1987) 1.


\bibitem{CDL1988}O. Casta{\~n}os, J. P. Draayer, Y. Leschber, Z. Phys. A \textbf{329}, (1988) 33.

\bibitem{Naqvi1995}H. A. Naqvi, C. Bahri, D. Troltenier, J. P. Draayer, A. Faessler, Z. Phys. A \textbf{351}, (1995) 259.

\bibitem{Smirnov2000}Yuri F. Smirnov, Nadya A. Smirnova, Piet Van Isacker, Phys. Rev. C \textbf{61}, (2000) 041302(R).

\bibitem{Thiamova2010}G. Thiamova, Eur. Phys. J. A \textbf{45}, (2010) 81.

\bibitem{Zhang2014}Y. Zhang, F. Pan, Lian-Rong Dai, and J. P. Draayer, Phys. Rev. C \textbf{90}, (2014) 044310.

\bibitem{Wood2004}J. L. Wood, A-M. Oros-Peusquens, R.Zaballa, J. M. Allmond, W. D. Kulp, Phys. Rev. C \textbf{70},
                  (2004) 024308; J. M. Allmond, Ph.D. dissertation, Georgia Insitute
                  of Technology(2007), http://hdl.handle.net/1853/14604.

\bibitem{Allmond2008}J. M. Allmond,  R. Zaballa, A-M. Oros-Peusquens, W. D. Kulp, J. L. Wood, Phys. Rev. C \textbf{78}, (2008)
014302.

\bibitem{Allmond2009}J. M. Allmond, J. L. Wood, W. D. Kulp, Phys. Rev. C \textbf{80}, (2009) 021303(R).

\bibitem{Allmond2010}J. M. Allmond, J. L. Wood, W. D. Kulp, Phys. Rev. C \textbf{81}, (2010) 051305(R).

\bibitem{Chen2014}Q. B. Chen, S. Q. Zhang, P. W. Zhao, J. Meng, Phys. Rev. C \textbf{90}, (2014) 044306.

\bibitem{Bonatsos2004}D. Bonatsos, D. Lenis, D. Petrellis, P. A. Terziev, Phys. Lett. B \textbf{588}, (2004) 172.

\bibitem{Zamfir1991}N. V. Zamfir, R. F. Casten, Phys. Lett. B \textbf{260}, (1991) 265.

\end{thebibliography}
\end{document}